\documentclass[onecolumn,aps,pra,nofootinbib]{revtex4-1}
\usepackage[utf8]{inputenc}
\usepackage{graphicx}
\usepackage{float}
\usepackage{mathtools}
\usepackage{quantikz}
\begin{document}

\title{Mixed-encoding one-photon-interference quantum secure direct communication}

\author{Xiang-Jie Li}
\affiliation{State Key Laboratory of Low-dimensional Quantum Physics and Department of Physics, Tsinghua University, Beijing 100084, China}

\author{Yuan-Bin Cheng}
\affiliation{State Key Laboratory of Low-dimensional Quantum Physics and Department of Physics, Tsinghua University, Beijing 100084, China}

\author{Xing-Bo Pan}
\affiliation{State Key Laboratory of Low-dimensional Quantum Physics and Department of Physics, Tsinghua University, Beijing 100084, China}

\author{Yun-Rong Zhang}
\affiliation{State Key Laboratory of Low-dimensional Quantum Physics and Department of Physics, Tsinghua University, Beijing 100084, China}

\author{Gui-Lu Long}
\email{gllong@mail.tsinghua.edu.cn}
\affiliation{State Key Laboratory of Low-dimensional Quantum Physics and Department of Physics, Tsinghua University, Beijing 100084, China}
\affiliation{Beijing Academy of Quantum Information Sciences, Beijing 100193, China}
\affiliation{Frontier Science Center for Quantum Information, Beijing 100084, China}
\affiliation{Beijing National Research Center for Information Science and Technology, Beijing 100084, China}

\date{\today}

\begin{abstract}
Quantum secure direct communication (QSDC) guarantees both the security and reliability of information transmission using quantum states. One-photon-interference QSDC (OPI-QSDC) is a technique that enhances the transmission distance and ensures secure point-to-point information transmission, but it requires complex phase locking technology. This paper proposes a mixed-encoding one-photon-interference QSDC (MO-QSDC) protocol that removes the need for phase locking technology. Numerical simulations demonstrate that the MO-QSDC protocol could also beat the PLOB bound.
\end{abstract}

\maketitle

\section{\label{sec:Introduction}Introduction\protect\\ }
Quantum computers pose a serious threat to classical encryption algorithms~\cite{shor1999polynomial,grover1997quantum,yan2022factoring,wang2022variational}, which has led to the development of quantum communication~\cite{BennettIEEE1984,LL00,DL04} to ensure secure information transmission. Quantum secure direct communication (QSDC)~\cite{LL00,DL04,Wang2005a,Wang2005b} is a significant paradigm in quantum communication that allows for the direct transmission of information using quantum states in the channel, providing the ability to detect and prevent eavesdropping.

Despite the theoretical proof of its unconditional security~\cite{qi2019,wu2019security,ye2021generic} and feasibility demonstrated through experiments~\cite{hu16experimental,zhang17experimental,zhu17experimental,zhang2022realization}, actual physical devices used in quantum communication systems have been found to have numerous security loopholes~\cite{huang2018implementation}, particularly attacks against measurement devices~\cite{qi2005time,makarov2005faked,makarov2009controlling}. To address these issues, measurement-device-independent QSDC (MDI-QSDC)~\cite{zhou2020measurement,niu2018measurement} protocols have been proposed, where encoded signals are sent to an untrusted third party, Charlie, for measurement and publication of the results. Alice and Bob can then determine whether the information is secure based on the results published by Charlie and proceed with further information exchange.

One-photon-interference QSDC (OPI-QSDC)~\cite{li2023one} protocol has been proposed to overcome practical application challenges of MDI-QSDC protocols, such as dependence on immature quantum memories and ideal single-photon sources, short transmission distances, and low secrecy rates. The information is encoded on the global phase of a coherent state in this protocol. Alice and Bob use techniques such as phase locking and phase tracking to lock the frequency and global phase of their lasers, allowing their signals to form single-photon interference~\cite{lucamarini2018overcoming,ma2018phase,curty2019simple,cui2019twin} at Charlie's location. Coherent states transmitted in optical fibers undergo phase evolution resulting in errors in the transmitted information. Phase-locking technology~\cite{lucamarini2018overcoming} can ensure that the bit information encoded in the global phase of the coherent state is not lost during transmission. The channel loss of the protocol is significantly reduced, being only one-fourth of that in the original MDI-QSDC protocol. This improvement enables the protocol to beat the PLOB bound, which is the repeaterless quantum communication rate limit~\cite{pirandola2017fundamental}.

However, the implementation of phase-locking technology for quantum communication can be challenging due to the complexity of fiber structures and auxiliary systems required~\cite{Wang2022}. To remove it and maintain low channel loss when using single-photon interference, we propose a mixed-encoding one-photon-interference QSDC (MO-QSDC) protocol by combining two approaches: intensity encoding and relative phase encoding instead of global phase encoding. In the intensity encoding approach~\cite{wang2018twin,zeng2022mode,xie2022breaking}, only Alice or Bob sends a photon, and the presence or absence of the photon is used to encode information. In the relative phase encoding approach~\cite{zeng2022mode,xie2022breaking}, two pulses emitted by the same laser with a short interval are used, which have the same phase fluctuation. As they pass through the same fiber and reach Charlie, the relative phase can be maintained stably without phase locking, thus ensuring that the bit information is not lost. However, when the interval between the two pulses is long, the relative phase cannot be determined, resulting in a lower secrecy rate~\cite{zhu2023experimental} or requiring a high quality light source~\cite{zhou2023experimental}. 
Our protocol determines the basis preparation strategy based on the information to be transmitted, which is secret for Eve beforehand, thus eliminating the effect of laser phase fluctuations on the secrecy rate. Our proposed protocol has been demonstrated through numerical simulations to beat the PLOB bound and may have the potential to play a significant role in quantum networks~\cite{long2022evolutionary}, free-space QSDC~\cite{pan2020experimental}, and other fields.

The remainder of this paper is organized as follows. In Sec.~\ref{sec:protocol} we describe the detailed steps of the proposed MO-QSDC protocol, while in Sec.~\ref{sec:security} we analyze its security. Then in Sec.~\ref{sec:performance} we present our numerical analysis of performance. Finally, we give a conclusion in Sec.~\ref{sec:conclusion}. 

\section{\label{sec:protocol}Protocol description\protect\\ }
\begin{figure*}
  \centering
   \includegraphics[width=\textwidth]{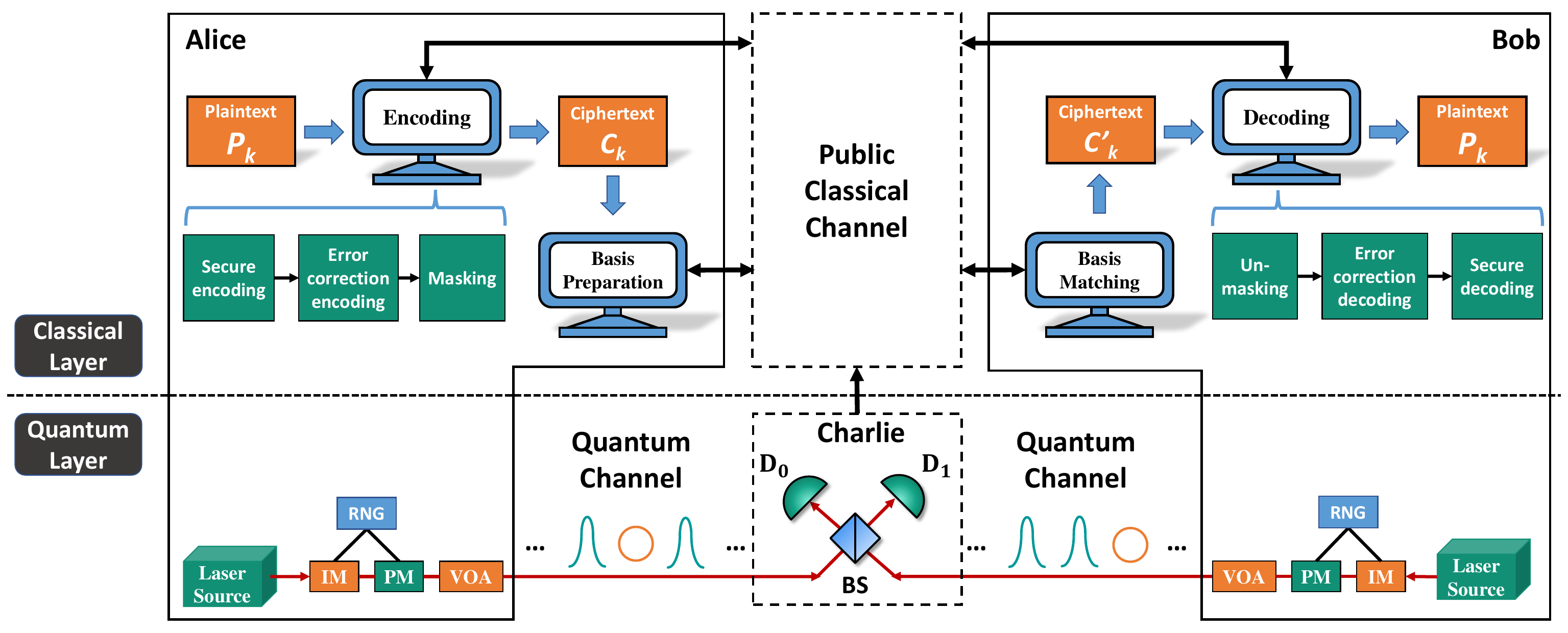}
   \caption{\label{fig:diagram}\textbf{Schematic diagram of MO-QSDC protocol in the $k$-th frame transmission.} BS: 50:50 beam splitter; $\rm D_0$, $\rm D_1$: single-photon detector; IM: intensity modulator; PM: phase modulator; RNG: random number generator; VOA: variable optical attenuator.}
\end{figure*}
To ensure the security of QSDC, the channel security must be confirmed before loading any information~\cite{LL00,DL04,li2023single}. Because high-performance quantum memory is not available at the time of writing, delayed secure coding can be used to guarantee the transmission security of information~\cite{qi2019,sun2020qmf,zhang2022realization}. This method involves dividing the pending transmission information into several frames, each with multiple bits. The result of the parameter estimation in the $(k-1)$-th frame or earlier frames ensures the security of the $k$-th frame information. For the first frame, random numbers can be transmitted to complete the parameter estimation.

When transmitting the $k$-th frame, both Alice and Bob simultaneously send signal pulses to Charlie, an untrusted third party located in the middle of them, as shown in the quantum layer of Fig.~\ref{fig:diagram}. While Alice performs error correction coding, secure encoding,~\cite{zhang2022realization} and masking~\cite{long2021drastic} locally to encode the plaintext $P_k$ into ciphertext $C_k$, as shown in the classical layer of Fig.~\ref{fig:diagram}. After Alice collects enough valid events, she assigns these events as the X-basis, Z-basis and O, and then publish them. The protocol's detailed steps of the $k$-th frame transmission are described as follows.

Firstly, Alice, Bob, and Charlie perform the following steps in the \textbf{quantum layer}:

\textbf{Step 1: State Preparation.} Alice prepares state $\ket{\alpha_j e^{i\phi^j_a}}$ and sends it to Charlie in the $j$-th($j = 1, 2, .., NT$) round of communication. $\alpha$ is randomly chosen from $\{0, \sqrt{\nu}, \sqrt{\mu}\}$, and $\phi\in[0,2\pi)$ is a random phase. Similarly, Bob prepares state $\ket{\beta_j e^{i\phi_b^j}}$ and sends it to Charlie.

\textbf{Step 2: Measurement.} Charlie performs the single-photon-interference measurement on the photons sent by Alice and Bob, and announces the measurement result. The measurement result is the clicks of two single-photon detectors, which we denote by $\rm D_0$ and $\rm D_1$($\rm D_0, \rm D_1 \in \{0, 1\}$). Here we use $0$ for no click events and $1$ for click events. Alice and Bob keep the state preparation information for the case $\rm D_0+\rm D_1=1$.

They repeat the above process $NT$ times, and then perform the following steps in the \textbf{classical layer}:

\textbf{Step 3: Encoding.} Alice encodes the plaintext $P_k$ using secure coding, error correction coding~\cite{zhang2022realization}, and masking~\cite{long2021drastic} to obtain the ciphertext $C_k$. For more information about the encoding process, please refer to Appendix~\ref{app:encoding}.

\textbf{Step 4: Basis Preparation.} To ensure the security of our protocol, Alice and Bob first verify the occurrence of events with no other detection events within a maximum interval of $T$, where $T$ is the unpublished maximum events interval that has the same phase evolution. They proceed with the protocol only if the number of such events is below a threshold $\Lambda $.

Next, Alice prepares the X bases, whose amplitudes are $(\alpha_i, \alpha_j)\in{(\sqrt{\mu_i}\backslash\sqrt{\nu_i},\sqrt{\mu_j}\backslash\sqrt{\nu_j})}$ for the $i$-th round and $j$-th round pulses, with a distance less than $T$.

Then, Alice prepares the Z bases. She pairs the remaining pulses in order according to the ciphertext $C_k$. The pairs for bit 0 are $(0_i, \sqrt{\mu_j}\backslash \sqrt{\nu_j})$ and for bit 1 are $( \sqrt{\mu_i}\backslash \sqrt{\nu_i},0_j)$. The relative phase of these two pulses in the Z-basis does not affect the information encoded by Alice, so the distance between them can be chosen arbitrarily.

Finally, Alice prepares the O, whose amplitudes are $(0_i, 0_j)$. They will be used for parameter estimation.

\begin{figure}
  \centering
   \includegraphics[width=\columnwidth]{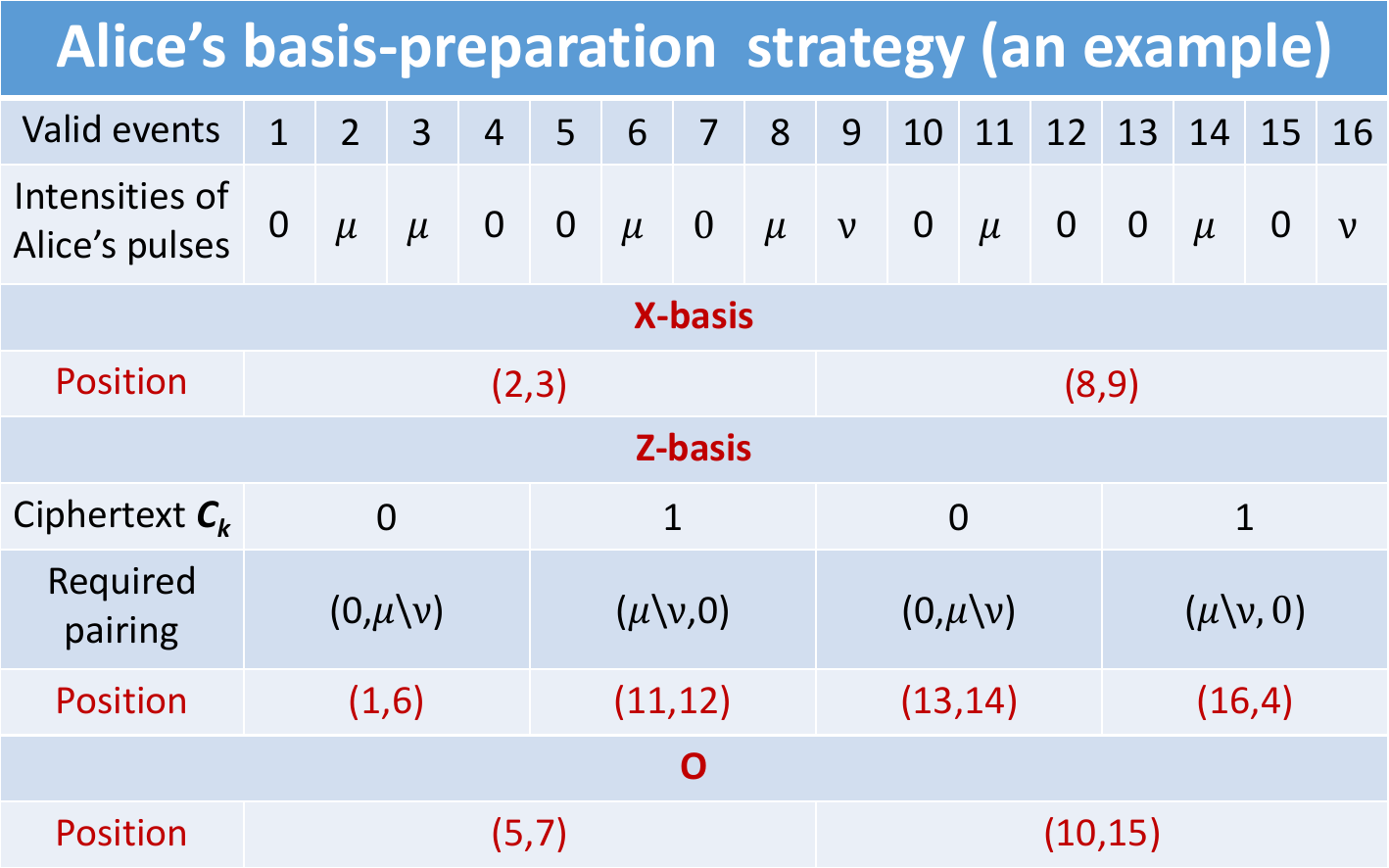}
   \caption{\label{fig:BP}\textbf{A simple example of Alice's basis-preparation strategy.} In this example, Alice intends to send a ciphertext of 0101. She has a total of 16 valid events, and she first assigns 4 of them as the X-basis. Then, she assigns 8 events as the Z-basis based on the ciphertext $C_k$. Finally, she assigns the remaining 4 events as O. The information that Alice needs to publish is highlighted in red. }
\end{figure}

In Fig.~\ref{fig:BP}, we provide a simple example to help readers better understand the basis preparation process.

\textbf{Step 5: Basis Matching.} Alice publishes the basis-preparation strategy. Alice and Bob discard the bases that are different. Note that there is a probability of $1-k \approx 7/8$ that the basis matching will fail, resulting in $C'_k \neq C_k$. However, Bob can use the error correction coding to recover the information.

\textbf{Step 6: Bit Mapping.} Alice and Bob map the matched bases to bits.

For Z bases, Bob maps $(0_i, \sqrt{\mu_j} \backslash \sqrt{\nu_j})$ to bit 1 and $( \sqrt{\mu_i}\backslash\sqrt{\nu_i},0_j)$ to bit 0.

For X bases, Alice and Bob extract the bit information from the relative phase of the two pulses. They compute $\phi^j_a -\phi^i_a = \theta_a + \pi b_a$ and $\phi^j_b -\phi^i_b = \theta_b + \pi b_b$, respectively, where $\theta_a, \theta_b \in [0, \pi)$. Then, they publish the intensities and relative phases $\theta_a, \theta_b$ of two pulses, and they keep the bases whose intensities and relative phases are matched and discard others. Alice uses $b_a$ as the bit information. Bob maps the bit information based on Charlie's measurement results. If the measurement result is $(0^i1^i, 1^j0^j)$ or $(1^i0^i, 0^j1^j)$, then Bob flips his bit $b_b$; otherwise, Bob keeps his bit unchanged.

\textbf{Step 7: Parameter Estimation.} Alice and Bob can obtain the average gain $Q_k$ and the quantum bit error rate (QBER) $E^Z_k$ from the experiment. They can then estimate the average single-photon bases rate $\Delta _{1,k}$ and the single photon phase error rate $e^X_{11,k}$, and calculate the secrecy rate $R_k$.

\textbf{Step 8: Decoding.} Based on the results of parameter estimation and the information Alice published, Bob decodes the ciphertext $C'_k$ using unmasking~\cite{long2021drastic}, error correction decoding and secure decoding~\cite{zhang2022realization} to obtain the plaintext $P_k$. For more information about the decoding process, please refer to Appendix~\ref{app:encoding}.

\section{\label{sec:security}Security analysis\protect\\ }
According to the quantum wiretap channel model~\cite{qi2019,wu2019security,ye2021generic}, the secrecy channel capacity is given by
\begin{equation}
  \begin{aligned}
    C_S&=C_M-C_W\\
    &=\max \limits_{P_X} [I(A:B)-I(A:E)],
  \end{aligned}
\end{equation}
where $P_X$ represents the probability distribution of information, while $C_M$ and $C_W$ represent the main channel capacity and the wiretap channel capacity, respectively; $I(A:B)$ and $I(A:E)$ represent the mutual information of Alice to Bob and Alice to Eve, respectively. Based on this, we can obtain an achievable secrecy rate $R$  to ensure secure and reliable transmission of information, where $0\leq R\leq C_S$ and
\begin{equation}
  R = I(A:B) - I(A:E).
\end{equation}
Based on channel model,
\begin{equation}
  I(A:B) = Q[1-h(E^Z)],
\end{equation} 
where $h(x)=-x\log (x)-(1-x)\log (1-x)$ is the binary entropy function, while $Q$ and $E^Z$ represent the average gain and QBER, respectively. For convenience, we omitted the subscript $k$ representing the $k$-th frame. We analyzed the eavesdropping process to calculate $I(A:E)$, as detailed in Appendix~\ref{app:security}. Then we obtained the achievable secure rate $R$ as 
\begin{equation}
    \label{equ:R}
  \begin{aligned}
    R = p\cdot q\cdot Q \{ \Delta_1 [1-h(e^X_{11})]-fh(E^Z)\},
  \end{aligned}
\end{equation}
where $p$ and $q$ represent the basis preparation rate and basis matching rate, respectively. $f$ represents the inefficiency function of error correction coding, while $\Delta _{1}$ and $e^X_{11}$ represent the average single-photon bases rate and the single-photon phase error rate, respectively.

\section{\label{sec:performance}Performance analysis\protect\\ }
\subsection{The achievable secrecy rate vs. transmission distance}
\begin{figure}[h]
    \centering
     \includegraphics[width=\columnwidth]{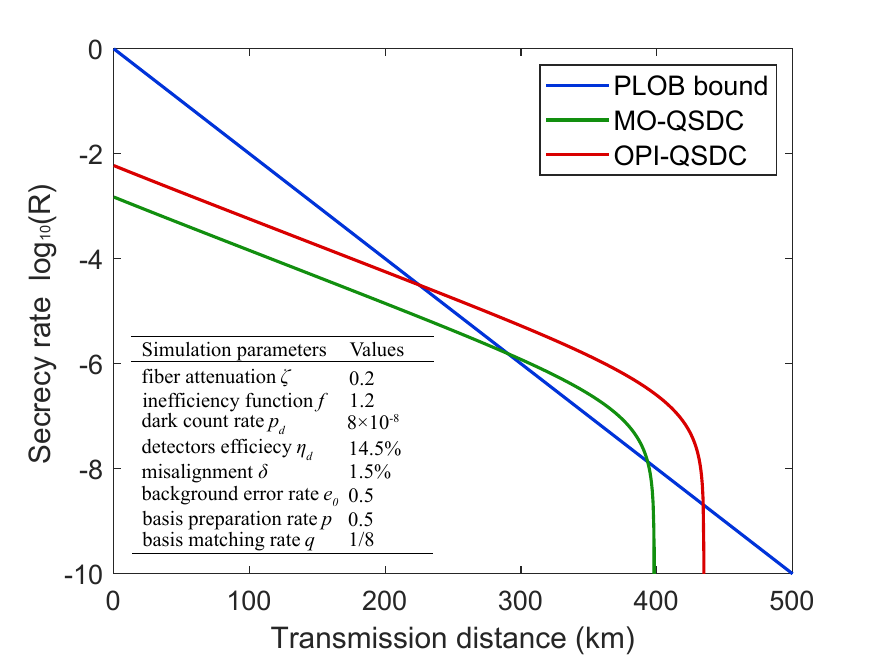}
     \caption{\label{fig:comparison}\textbf{The achievable secrecy rate $log_{10}(R)$ vs. transmission distance.} The blue line represents the PLOB pound, while the green and red lines represent our parameterized MO-QSDC protocol and OPI-QSDC protocol, respectively. The simulation parameters are listed in the table.}
\end{figure}
We simulated the performance of our MO-QSDC protocol in Fig.\ref{fig:comparison}. The detailed simulation formulas can be found in Appendix \ref{app:performance}. Since the MO-QSDC protocol uses only single-photon components for encoding, while the OPI-QSDC protocol allows for multi-photon components, the MO-QSDC protocol has a lower secrecy rate and shorter transmission distance compared to the OPI-QSDC protocol, as shown in Fig.\ref{fig:comparison}. However, the MO-QSDC protocol can still surpass the PLOB bound\cite{pirandola2017fundamental} for distances $d > 290 \rm km$, with a maximum transmission distance of $400 \rm km$.

\subsection{The success rate of X-basis preparation}
\begin{figure}[h]
    \centering
     \includegraphics[width=\columnwidth]{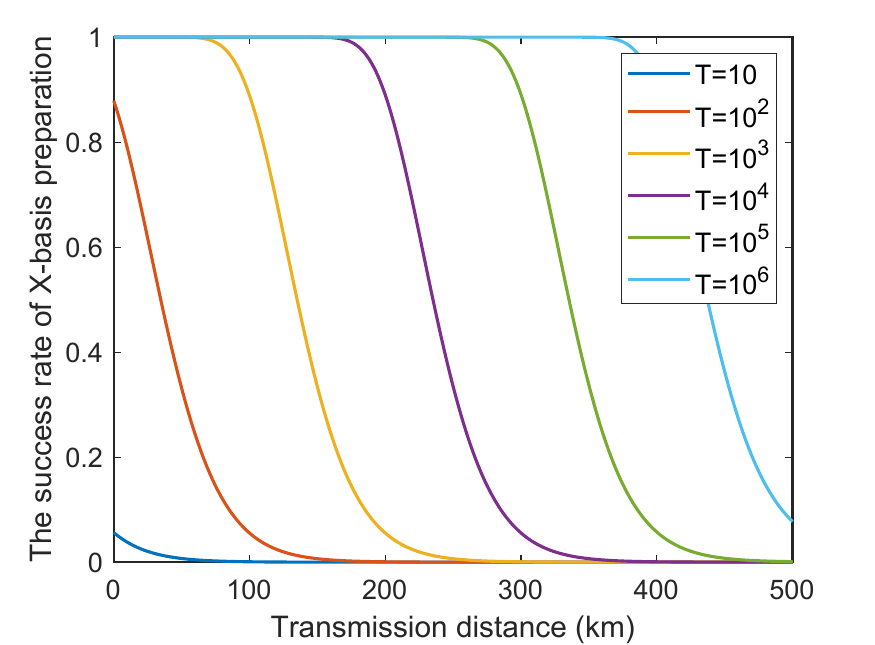}
     \caption{\label{fig:T}\textbf{The success rate of X-basis preparation vs. transmission distance when there are only $T$ events.} The lines of different colors represent different values of the maximum event interval $T$.}
\end{figure}
In \textbf{Step 4} of our MO-QSDC protocol, Alice needs to use a relative phase to prepare the X-basis. However, when the interval between two events exceeds $T$, the relative phase of the two events cannot be determined. As a result, it becomes impossible to encode the X-basis using those two events. When the total number of events sent is limited to $T$, there is a probability that no two events are suitable for preparing the X-basis. Therefore, the success rate $Pr(X|T)$ of X-basis preparation when the total number of events sent is $T$ can be written as
\begin{equation}
    \begin{aligned}
        Pr(X|T)=&\mathcal{C}_T^2 Pr^2(E_{\mu},E_{SPI})[1-Pr(E_{\mu},E_{SPI})]^{T-2}\\
        &+\mathcal{C}_T^3 Pr^3(E_{\mu},E_{SPI})[1-Pr(E_{\mu},E_{SPI})]^{T-3}\\
        &+...+\mathcal{C}_T^T Pr^T(E_{\mu},E_{SPI})\\
      =&1-\mathcal{C}_T^1 Pr(E_{\mu},E_{SPI})[1-Pr(E_{\mu},E_{SPI})]^{T-1}\\
      &-\mathcal{C}_T^0 [1-Pr(E_{\mu},E_{SPI})]^T,
    \end{aligned}
\end{equation}
where $\mathcal{C}_m^n=m!/n!/(m-n)! $ is the binomial coefficient, and $Pr(E_{\mu}, E_{SPI})$ represents the probability that the event is a single-photon-interference event and the intensity of the pulse sent by Alice is $\mu$. We assume that Alice and Bob send pulses of $0$ intensity with a probability of $1/2$, and pulses of $\mu$ intensity with a probability of $1/2$. Thus, we have $Pr(E_{\mu},E_{SPI})=Q/2$. The success rate of X-basis preparation vs. transmission distance when there are only $T$ events is shown in Fig.~\Ref{fig:T}. We set $z=(T-1)\cdot Pr(E_{\mu},E_{SPI})$ and make a further approximation to $Pr(X|T)$, that is,
\begin{equation}
        Pr(X|T)\approx 1-(1+z)e^ {-z}.
\end{equation}
We have
\begin{equation}
    \lim_{z \to 0}  Pr(X|T)=0.
\end{equation}
Hence, as the transmission distance increases, the maximum event interval $T$ required also increases to ensure the success rate of X-basis preparation.

In general, the length of the maximum event interval $T$ depends on the phase fluctuations of lasers. However, when $T$ is small, we can increase the number of events sent to ensure the success rate of X-basis preparation, i.e., we send $NT$ events per frame. We define $Pr(F|NT)$ as the failure rate of X-basis preparation when there are $NT$ events, then we have
\begin{equation}
    \label{equ:NT}
    Pr(F|NT)=[1-Pr(X|T)]^N.
\end{equation}
We notice that
\begin{equation}
    \lim_{N \to \infty }Pr(F|NT)=0,
\end{equation}
thus when a sufficient number of events are sent per frame, the success rate of X-basis preparation can always be ensured.

\begin{figure}[h]
    \centering
     \includegraphics[width=\columnwidth]{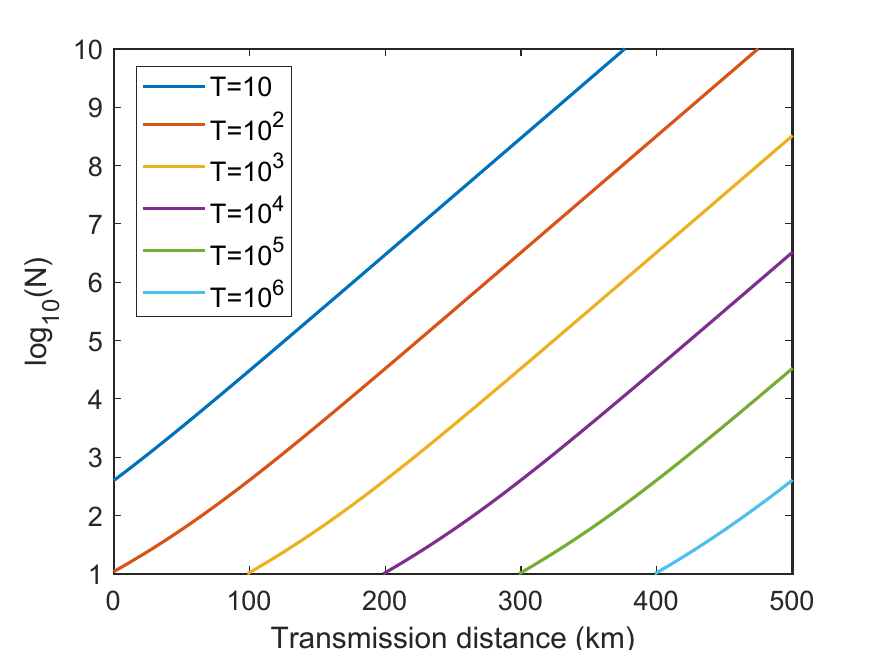}
     \caption{\label{fig:N}\textbf{$\log_{10}N$ vs. transmission distance when the failure rate of X-basis preparation is $10^{-10}$.} The lines of different colors represent different values of the maximum event interval $T$. $N$ is the required number to ensure the X-basis preparation failure rate is $10^{-10}$ when there are $NT$ events.}
\end{figure}

Equ. (\ref{equ:NT}) can be written as
\begin{equation}
    N=\frac{\log_{10}Pr(F|NT)}{\log_{10}[1-Pr(F|NT)]}.
\end{equation}
We set the failure rate $Pr(F|NT)$ to $10^{-10}$, and then plotted the required number $N$ vs. transmission distance, as shown in Fig.~\ref{fig:N}. For convenience, we used $\log_{10}N$ as the vertical axis.

Lastly, it is important to note that in our MO-QSDC protocol, we use the Z-basis to transmit information and use the X-basis to estimate the amount of information leakage. A small number of X-basis are sufficient to accurately estimate the leakage~\cite{Lo2005a}, which means that the number of X-basis required is much smaller than the Z-basis. Therefore, when a sufficient number of events are sent, the effect of the event interval $T$ on the secrecy rate can be disregarded.

\section{\label{sec:conclusion}Conclusions\protect\\ }
In summary, we proposed MO-QSDC protocol and analyzed its security using quantum wiretap channel theory. The performance analysis shows that our protocol could achieve a maximum transmission distance of approximately 400 km when using non-ultra-low loss optical fiber, while the secrecy rate remains immune to laser source phase fluctuations when transmitting sufficiently long information. Additionally, it has the capability to surpass the PLOB bound beyond approximately 290 km. Although, its performance may be slightly less impressive than that of the OPI-QSDC protocol, it has the potential to provide a more robust and efficient solution for QSDC, particularly in situations where phase-locking technology is difficult to implement.

\begin{acknowledgements}
This work is supported by National Key R\&D Program of China (2017YFA0303700), the Key R\&D Program of Guangdong province (2018B030325002), Beijing Advanced Innovation Center for Future Chip (ICFC), Tsinghua University Initiative Scientific Research Program and the National Natural Science Foundation of China under Grants No. 61727801, No. 11974205, and No. 11774197. 

\end{acknowledgements}

\clearpage

\begin{widetext}
\appendix
\section{Details of the encoding and decoding process}
\label{app:encoding}
\begin{figure*}[h]
  \centering
   \includegraphics[width=\textwidth]{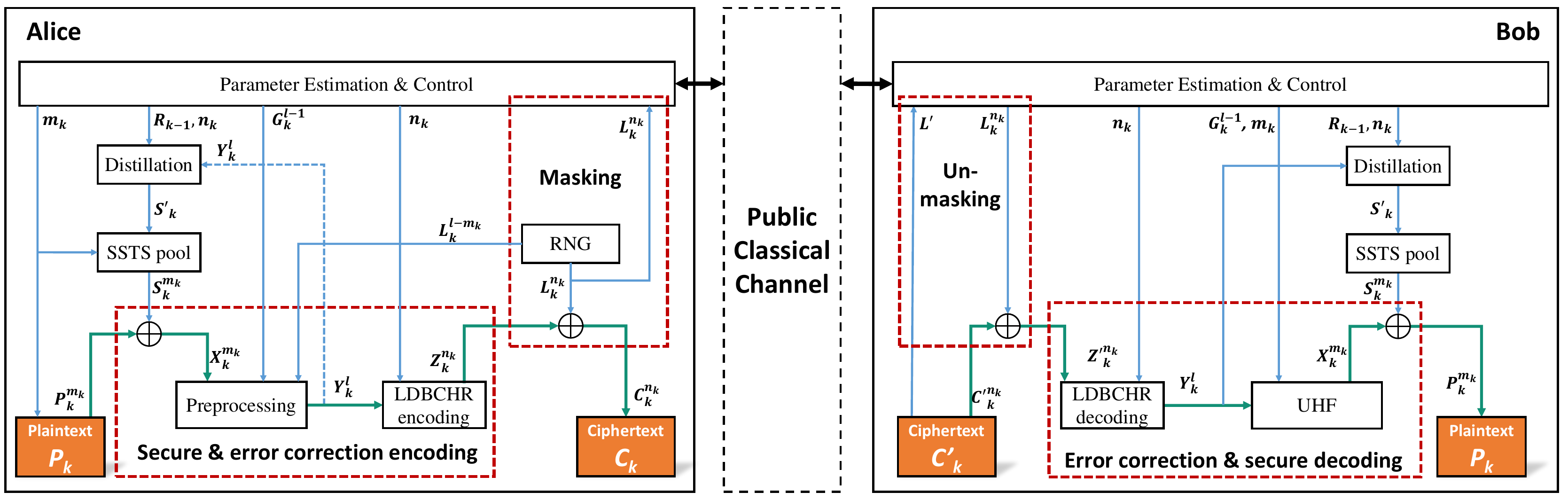}
   \caption{\label{fig:encoding}\textbf{Detailed flowchart of the $k$-th frame encoding and decoding.} SSTS: shared secure transmission sequence; LDBCH: low-density Bose-Chaudhuri-Hocquenghem concatenated with repetition; RNG: random number generator; UHF: universal hashing families. The green arrows indicate the flow of information data blocks, and the blue arrows indicate the auxiliary data flow. The meaning of symbols are shown in Table.~\ref{tab:encoding}.}
\end{figure*}
\begin{table}[h]
  \centering
  \caption{\label{tab:encoding}The meaning of symbols in Fig.\ref{fig:encoding}.}
  \begin{tabular}{cc}
  \hline 
  \hline
Symbol           & Description  \\ \hline 
$m_k$            & The length of the plaintext to be transmitted in the $k$-th frame. \\
$P_k^{m_k}$      & $m_k$ bits plaintext.\\
$S_k^{m_k}$      & The SSTS used to encrypt $P_k^{m_k}$. \\
$X_k^{m_k}$      & The intermediate ciphertext encrypted by $S_k^{m_k}$, $X_k^{m_k}=S_k^{m_k}\oplus P_k^{m_k}$.\\
$l$              & The length of the cipertext after secure coding.\\
$L_{k}^{l-m_k},L_{k}^{n_k}$           & The local random number sequence.\\
$G_{k}^{l-1}$                      & The global random number sequence that will be published subsequently.\\
$Y_k^l$          & The output sequence after reverse hash mapping, with a length of $l$.\\
$n_k$            & The length of the cipertext after error correction coding.\\
$Z_k^{n_k}$      & The intermediate ciphertext after LDBCHR $(n_k, l)$ error correction coding with a length of $n_k$.\\
$C_k^{n_k}$      & The final ciphertext encrypted by $L_{k}^{n_k}$, $C_k^{n_k}=L_{n_k}^{n_k}\oplus Z_k^{n_k}$.\\
$R_{k-1}$        & The achievable secrecy rate of the $(k-1)$-th frame.\\
$S'_k$           & The SSTS shared by Alice and Bob distilled from $X_k^l$.\\
$L'$             & The positions where Bob received the information.\\
\hline
\hline
\end{tabular}
\end{table}
The flowchart depicts the encoding and decoding process of the $k$-th frame, as shown in Fig~\ref{fig:encoding}. The secure coding is the preprocessing procedure, which mainly involves Alice and Bob using the same shared secure transmission sequence (SSTS) to encrypt information and performing the inverse hash function mapping. The error correction coding is a low-density Bose-Chaudhuri-Hocquenghem concatenated with repetition (LDBCHR) coding proposed in~\cite{zhang2022realization}.

Assuming that the achievable secrecy rate $R_{k-1}$ for the $(k-1)$-th frame, as well as $I_{k-1}(A:B)$ and $I_{k-1}(A:E)$, have already been obtained before transmitting the $k$-th frame, and the encoding process must satisfy the following conditions based on the security analysis in Sec.~\ref{sec:security}:
\begin{equation}
\begin{aligned}
\frac{l}{n_k}&\leq  I_{k-1}(A:B), \\
\frac{m_k}{n_k}&\leq \frac{l}{n_k}-I_{k-1}(A:E)\\
& \leq R_{k-1}.
\end{aligned}
\end{equation}
These conditions ensure that the error correcting coding rate is less than the mutual information of the main channel, and the secure coding rate is less than the achievable secrecy rate of the channel, thereby ensuring the reliability and security of the communication, respectively.

We first describe the \textbf{Step 3: Encoding} process of our MO-QSDC protocol in detail below:

1. Alice encrypts the plaintext $P_k^{m_k}$ using SSTS $S_k^{m_k}$ through an XOR operation to obtain the intermediate ciphertext $X_k^{m_k}$, $X_k^{m_k}=S_k^{m_k}\oplus P_k^{m_k}$.

2. She performs the preprocessing step: $(X_k^{m_k},G_{k}^{l-1},L_{k}^{l-m_k})\stackrel{f^{-1}}{\longrightarrow}Y_k^l$, where $f^{-1}$ represents a family of reverse hash functions.

3. She uses LDBCHR encoding to transform $Y_k^l$ into $Z_k^{n_k}$.

4. She XORs the intermediate ciphertext $X_k^{m_k}$ with the local random number $L_{k}^{n_k}$ bit by bit to obtain the final ciphertext $C_k^{n_k}$, $C_k^{n_k}=L_k^{n_k}\oplus X_k^{m_k}$.

We then describe the \textbf{Step 8: Decoding} process of our MO-QSDC protocol in detail below:

1. Bob publishes the positions $L'$ where he received the information, and Alice publishes the random numbers $L_{k}^{n_k}$ at these positions, while the random numbers at the remaining positions will never be published. Bob XORs the ciphertext $C_k^{'n_k}$ with these random numbers to obtain $Z_k^{'n_k}$, $Z_k^{'n_k}=L_k^{n_k}\oplus C_k^{'n_k}$.

2. Bob uses LDBCHR decoding to transform $Z_k^{'n_k}$ into $Y_k^l$.

3. He performs the UHF step: $(Y_k^l,G_{k}^{l-1})\stackrel{f}{\longrightarrow}X_k^{m_k}$.

4. Finally, Alice and Bob distill the same SSTS $S'_k$ from $Y_k^l$ and insert it into the SSTS pool.

\section{Security analysis}
\label{app:security}
OPI-QSDC protocol~\cite{li2023one} using global phase encoding has the nature of resisting photon-number-splitting (PNS) attacks because of the uncertainty relationship between the global phase and the photon number. But this nature is not present when using relative phase encoding~\cite{zeng2022mode}. Therefore, in order to calculate I(A:E), we performed an equivalent entanglement transformation on the protocol. We first analyzed the achievable secrecy rate of the single-photon entanglement-based MO-QSDC protocol (Protocol \uppercase\expandafter{\romannumeral1}), then analyzed the achievable secrecy rate of the multi-photon entanglement-based MO-QSDC protocol (Protocol \uppercase\expandafter{\romannumeral2}), and finally obtained the achievable secrecy rate of our MO-QSDC protocol (Protocol \uppercase\expandafter{\romannumeral3}).

\subsection{Protocol \uppercase\expandafter{\romannumeral1}}
\begin{figure*}[h]
  \centering
   \includegraphics[width=\textwidth]{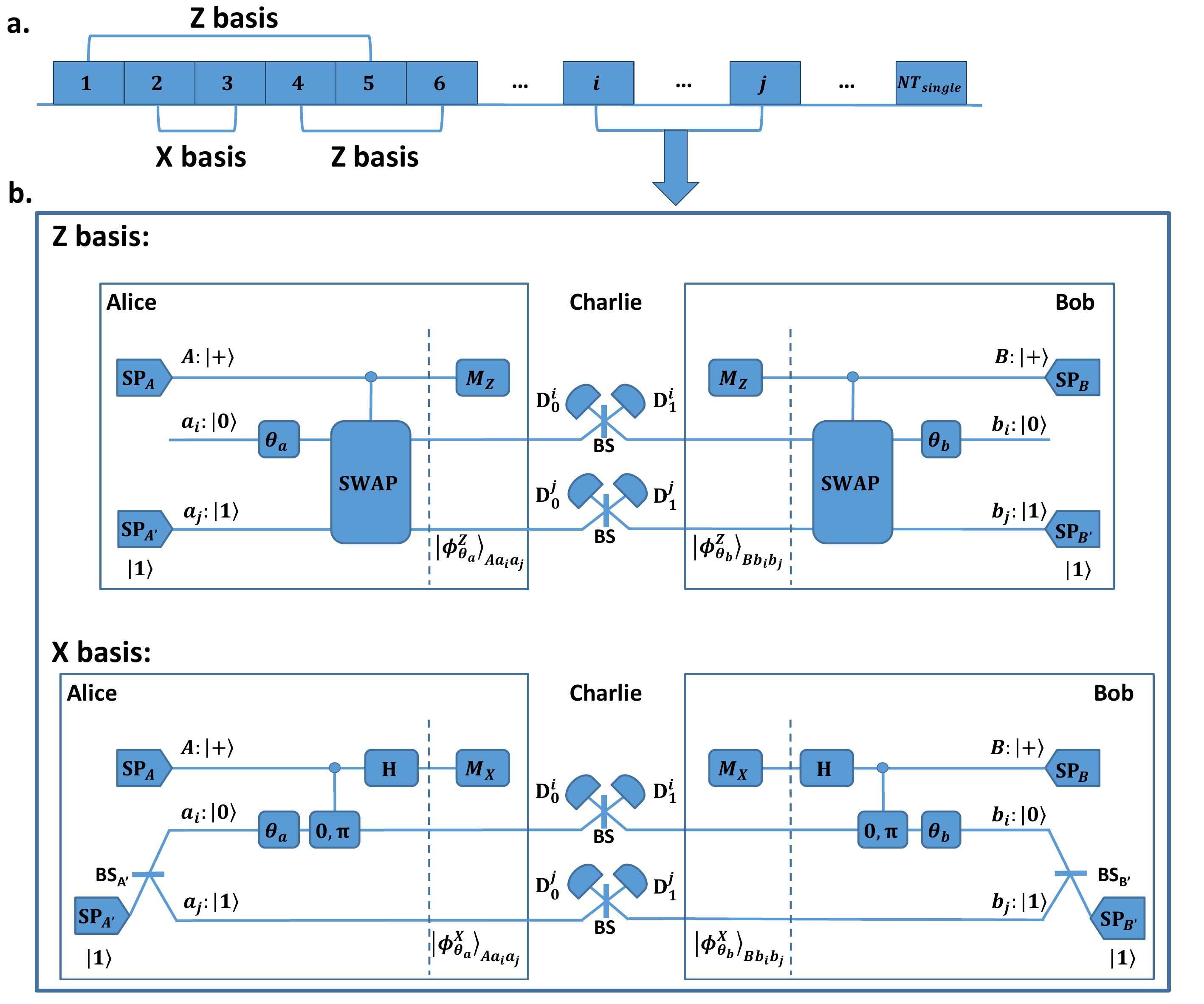}
   \caption{\label{fig:security_single}\textbf{Diagram of Protocol \uppercase\expandafter{\romannumeral1}.} BS, $\rm BS_{A'}$, $\rm BS_{B'}$: 50:50 beam splitter; $\rm D^i_0$, $\rm D^i_1$, $\rm D^j_0$ and $\rm D^j_1$: single-photon detector; $\rm SP_A$, $\rm SP_{A'}$, $\rm SP_B$ and $\rm SP_{B'}$: single-photon source. \textbf{a.} Schematic diagram of the basis preparation (an example). The numbers inside the boxes represent the serial numbers of the single-photon-interference events. \textbf{b.} Diagram of Protocol \uppercase\expandafter{\romannumeral1}'s quantum layer for pair $(i,j)$.}
\end{figure*}
The diagram of Protocol \uppercase\expandafter{\romannumeral1} is shown in Fig.~\ref{fig:security_single}. We omitted the diagram of the classical layer, because it has the same structure as Fig~\ref{fig:diagram}. The detailed steps of Protocol \uppercase\expandafter{\romannumeral1} are as follows.

\textbf{Step \uppercase\expandafter{\romannumeral1}.1: Encoding.} Same as Protocol \uppercase\expandafter{\romannumeral3}.3.

\textbf{Step \uppercase\expandafter{\romannumeral1}.2: Basis Preparation.} Alice and Bob jointly randomly pre-pair a total of $NT_{single}$ single-photon-interference events that will occur in the subsequent steps of this protocol, into $\left\lfloor NT_{single}/2 \right\rfloor $ pairs, through discussions by a secret channel, as shown in Fig.~\ref{fig:security_single}a. The distance of the two events in one pair is arbitrary. Next, Alice and Bob randomly choose a total of $N_X$ events to be in the X basis, respectively, where $N_X/(NT_{single})=p_x$. The distance of the two events in X basis should less than $T_{single}$. Then, they choose the remaining total of $N_Z$ pairs to be in the Z-basis, where $N_Z/(NT_{single})=p_z$, $p_x + p_z = 1$, and $p_x << p_z$.

\textbf{Step \uppercase\expandafter{\romannumeral1}.3: State Preparation.} We take the pair $(i,j)$ for example and denote the two pulses in this pair by $a_i$ and $a_j$ for Alice and $b_i$ and $b_j$ for Bob, as shown in Fig.~\ref{fig:security_single}b. If Alice or Bob chooses the Z basis, they send the quantum state 
\begin{equation}
    \begin{aligned}
        \ket{ \phi^{Z}_{\theta_a} }_{ A a_i a_j }&=\frac{1}{\sqrt{2}}(\ket{0}_A \ket{01}_{ a_i a_j}+ e^{i\theta_a} \ket{1}_A \ket{10}_{ a_i a_j }),\\
        \ket{ \phi^{Z}_{\theta_b} }_{ B b_i b_j }&=\frac{1}{\sqrt{2}}(\ket{0}_B \ket{01}_{ b_i b_j}+ e^{i\theta_b} \ket{1}_B \ket{10}_{ b_i b_j }).\\
    \end{aligned}
\end{equation}
If Alice or Bob chooses the X basis, they send the quantum state
\begin{equation}
    \begin{aligned}
        \ket{ \phi^{X}_{\theta_a} }_{ A a_i a_j }&=\frac{1}{\sqrt{2}}(\ket{+}_A \ket{\Psi _{\theta_a}^+}_{ a_i a_j}+ \ket{-}_A \ket{\Psi _{\theta_a}^-}_{ a_i a_j }),\\
        \ket{ \phi^{X}_{\theta_b} }_{ B b_i b_j }&=\frac{1}{\sqrt{2}}(\ket{+}_B \ket{\Psi _{\theta_b}^+}_{ b_i b_j}+ \ket{-}_B \ket{\Psi _{\theta_b}^-}_{ b_i b_j }),\\
    \end{aligned}
\end{equation}
where
\begin{equation}
    \begin{aligned}
        \ket{ \Psi _{\theta_a}^\pm  } &= \frac{1}{\sqrt{2}} ( \ket{01}\pm e^{i \theta_a} \ket{10}),\\
        \ket{ \Psi _{\theta_b}^\pm } &= \frac{1}{\sqrt{2}} ( \ket{01}\pm e^{i \theta_b} \ket{10}).\\
    \end{aligned}
\end{equation}

\textbf{Step \uppercase\expandafter{\romannumeral1}.4.1: Charlie's Measurement.} Same as Protocol \uppercase\expandafter{\romannumeral3}.2.

\textbf{Step \uppercase\expandafter{\romannumeral1}.4.2: Alice and Bob's Measurement.} Alice and Bob measure the local qubits $A$ and $B$. Here, we assume that Alice can perform a deterministic measurement based on the ciphertext in the Z basis. For example, if she wants to send a ciphertext of 0, the measurement result of her local bit $A$ will be $\ket{0}$ under the Z basis.

\textbf{Step \uppercase\expandafter{\romannumeral1}.5: Basis Matching.} Same as Protocol \uppercase\expandafter{\romannumeral3}.5.

\textbf{Step \uppercase\expandafter{\romannumeral1}.6: Bit Mapping.} Alice and Bob map the matched bases to bits.

For Z bases, it is the same as Protocol \uppercase\expandafter{\romannumeral3}.6.

For X bases, they only need to publish the relative phases $\theta_a, \theta_b$ of two pulses, and they keep the bases whose relative phases are matched and discard others. Alice maps the joint states of $a_i,a_j$ as follows: $\ket{\phi^{+}_{\theta_a} }_{ A a_i a_j} \rightarrow  0$, $\ket{\phi^{-}_{\theta_a} }_{ A a_i a_j} \rightarrow  1$. Bob maps the bit information based on Charlie's measurement results. If the measurement result is $(0^i1^i, 1^j0^j)$ or $(1^i0^i, 0^j1^j)$, Bob flips his bit; otherwise, Bob keeps his bit unchanged.

\textbf{Step \uppercase\expandafter{\romannumeral1}.7: Parameter Estimation.} Alice and Bob use the yield $Y_{11}$, QBER $e^Z_{11}$ and the phase error rate $e^X_{11}$ to calculate the achievable secrecy rate $R_{11}$.

\textbf{Step \uppercase\expandafter{\romannumeral1}.8: Decoding.} Same as Protocol \uppercase\expandafter{\romannumeral3}.8.

Next, we will analyze the steps of the protocol and calculate the achievable secrecy rate $R_{11}$. 

In \textbf{Step \uppercase\expandafter{\romannumeral1}.3}, we note that $\ket{ \phi _{\theta_a}^Z } = \ket{ \phi _{\theta_a}^X }$, $\ket{ \phi _{\theta_b}^Z } = \ket{ \phi _{\theta_b}^X }$ and Charlie who is under the full control of Eve, is unaware of Alice and Bob's secret basis preparation strategy at this moment. Hence, Charlie cannot distinguish them. We assume that Eve performs a collective attack using an auxiliary system $|E\rangle$ which can be generalized to coherent attacks using the quantum de Finetti theorem~\cite{renner2007symmetry}. After \textbf{Step \uppercase\expandafter{\romannumeral1}.4.1}, a purification state of Alice, Bob and Eve can be written as
\begin{equation}
    \ket{\phi}_{ABE} = \sum_{i=1}^4 \sqrt{\lambda_i} \ket{\Psi_i} \ket{E_i}
\end{equation}
where $\ket{\Psi_i}$ is one of the four Bell states: $\ket{\Psi^\pm}$, $\ket{\Phi^\pm}$, and $|E_1\rangle$, $|E_2\rangle$, $|E_3\rangle$ and $|E_4\rangle$ are the orthogonal states of Eve's auxiliary system $|E\rangle$. Assuming that in the case of without eavesdropping, Alice and Bob share the quantum state $\ket{\Psi}^+_{AB}$ perfectly. Then, the error rate in the X basis will be $e^X_{11} = \lambda_1 + \lambda_3$ and the error rate in the Z basis will be $e^Z_{11}=\lambda_3+\lambda_4$. Accordingly, the state of Alice and Eve is given by
\begin{equation}
    \begin{aligned}
        \rho_{AE} &= \text{Tr}_B(\ket{\phi}_{ABE}\bra{\phi})\\
            &= \frac{1}{2} (\braket{0}{_B \phi}_{ABE} \braket{\phi}{0}_B +\braket{1}{_B \phi}_{ABE} \braket{\phi}{1}_B  )\\
            &= \frac{1}{2} (\ket{\varphi}_1\bra{\varphi} + \ket{\varphi}_2\bra{\varphi}),
    \end{aligned}
\end{equation}
where we define 
\begin{equation}
    \begin{aligned}
        \ket{\varphi}_1 & \equiv \ket{0}_A(\sqrt{\lambda_{3}}\ket{E_3} + \sqrt{\lambda_{4}}\ket{E_4}) + \ket{1}_A (-\sqrt{\lambda_{1}} \ket{E_1} + \sqrt{\lambda_{2}} \ket{E_2}),\\
        \ket{\varphi}_2 & \equiv \ket{0}_A(\sqrt{\lambda_{1}}\ket{E_1} + \sqrt{\lambda_{2}}\ket{E_2}) + \ket{1}_A (-\sqrt{\lambda_{3}} \ket{E_3} + \sqrt{\lambda_{4}} \ket{E_4}).\\
    \end{aligned}
\end{equation}

In \textbf{Step \uppercase\expandafter{\romannumeral1}.4.2}, after Alice performs measurements, the states of Eve are given by
\begin{equation}
    \begin{aligned}
        \rho^0_E &= \text{Tr}_A(\ket{0}_A\bra{0} \rho_{AE} \ket{0}_A\bra{0}) \\
            &= \ket{\psi_1}\bra{\psi_1} + \ket{\psi_2}\bra{\psi_2},\\
        \rho^1_E &= \text{Tr}_A(\ket{1}_A\bra{1} \rho_{AE} \ket{1}_A\bra{1})\\
            &= \ket{\psi_3}\bra{\psi_3} + \ket{\psi_4}\bra{\psi_4},\\
    \end{aligned}
\end{equation}
where we define $\ket{\psi_1} \equiv \sqrt{\lambda_{3}}\ket{E_3} + \sqrt{\lambda_{4}}\ket{E_4}, \ket{\psi_2} \equiv \sqrt{\lambda_{2}}\ket{E_1} + \sqrt{\lambda_{2}} \ket{E_2}, \ket{\psi_3} \equiv -\sqrt{\lambda_{1}}\ket{E_1} + \sqrt{\lambda_{2}}\ket{E_2}, \ket{\psi_4}\equiv -\sqrt{\lambda_{3}}\ket{E_3} + \sqrt{\lambda_{4}}\ket{E_4}$.
Then, the mutual information between Alice and Eve is
\begin{equation}
    I(A:E)_{11} = S(\sum_k p_k \rho_E^k) -\sum_k p_k S(\rho_E^k)\leq h(e_{11}^X),
\end{equation}
where $k=0,1$, $p_k=1/2$.

Therefore we get the the achievable secrecy rate $R_{11}$, which is
\begin{equation}
        R_{11} = Y_{11} [1-h(e^Z_{11})-h(e^X_{11})].
\end{equation}

\subsection{Protocol \uppercase\expandafter{\romannumeral2}}
\begin{figure*}[h]
  \centering
   \includegraphics[width=\textwidth]{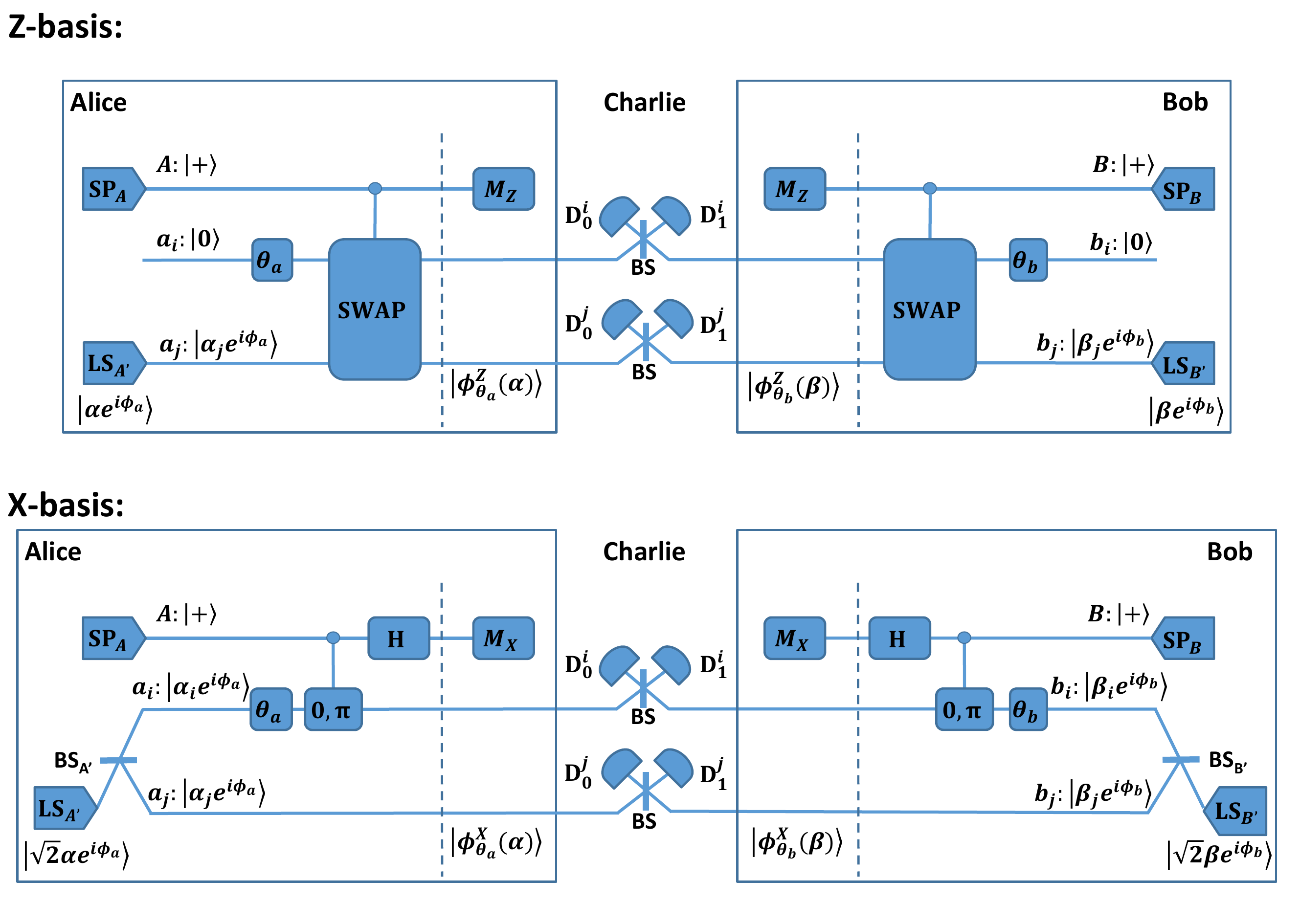}
   \caption{\label{fig:security_coherent}\textbf{Diagram of Protocol \uppercase\expandafter{\romannumeral2}'s quantum layer for pair $(i,j)$.} BS, $\rm BS_{A'}$, $\rm BS_{B'}$: 50:50 beam splitter; $\rm D^i_0$, $\rm D^i_1$, $\rm D^j_0$ and $\rm D^j_1$: single-photon detector; $\rm SP_A$, $\rm SP_{B}$: single-photon source; $\rm LS_A'$ and $\rm LS_{B'}$: laser source.}
\end{figure*}
The diagram of Protocol \uppercase\expandafter{\romannumeral2}'s quantum layer is shown in Fig.~\ref{fig:security_coherent}. The coherent light sources are replacing the single photon sources in $A'$ and $B'$. We also omitted the diagram of the classical layer and the detailed steps of Protocol \uppercase\expandafter{\romannumeral2} are as follows.

\textbf{Step \uppercase\expandafter{\romannumeral2}.1: Encoding.} Same as Protocol \uppercase\expandafter{\romannumeral3}.3.

\textbf{Step \uppercase\expandafter{\romannumeral2}.2: Basis Preparation.} Same as Protocol \uppercase\expandafter{\romannumeral1}.2.

\textbf{Step \uppercase\expandafter{\romannumeral2}.3: State Preparation.} If Alice or Bob chooses the Z basis, they send the quantum state 
\begin{equation}
    \begin{aligned}
        \ket{ \phi^{Z}_{\theta_a} }_{ A a_i a_j }&=\frac{1}{\sqrt{2}}(\ket{0}_A \ket{0}_{ a_i} \ket{\alpha e^{i\phi _a}}_{a_j}+ \ket{1}_A \ket{\alpha e^{i(\phi _a+\theta _a)}}_{a_i} \ket{0}_{a_j}),\\
        \ket{ \phi^{Z}_{\theta_b} }_{ B b_i b_j }&=\frac{1}{\sqrt{2}}(\ket{0}_B \ket{0}_{ b_i} \ket{\beta e^{i\phi _b}}_{b_j}+ \ket{1}_B \ket{\beta e^{i(\phi _b+\theta _b)}}_{b_i} \ket{0}_{b_j}),\\
    \end{aligned}
\end{equation}
where $\alpha$ and $\beta$ are randomly chosen from $\{ \sqrt{\nu} , \sqrt{\mu} \}$, and $\phi_a, \phi_b \in[0,2\pi)$ are random phases.
If Alice or Bob chooses the X basis, they send the quantum state
\begin{equation}
    \begin{aligned}
        \ket{ \phi^{X}_{\theta_a} }_{ A a_i a_j }&=\frac{1}{\sqrt{2}}(\ket{+}_A \ket{\alpha e^{i(\phi _a+\theta _a)}}_{ a_i} \ket{\alpha e^{i\phi _a}}_{a_j}+ \ket{-}_A \ket{-\alpha e^{i(\phi _a+\theta _a)}}_{a_i} \ket{\alpha e^{i\phi _a}}_{a_j}),\\
        \ket{ \phi^{X}_{\theta_b} }_{ B b_i b_j }&=\frac{1}{\sqrt{2}}(\ket{+}_B \ket{\beta e^{i(\phi _b+\theta _b)}}_{ b_i} \ket{\beta e^{i\phi _b}}_{b_j}+ \ket{-}_B \ket{-\beta e^{i(\phi _b+\theta _b)}}_{b_i} \ket{\beta e^{i\phi _b}}_{b_j}).\\
    \end{aligned}
\end{equation}

\textbf{Step \uppercase\expandafter{\romannumeral2}.4.1: Charlie's Measurement.} Same as Protocol \uppercase\expandafter{\romannumeral3}.2.

\textbf{Step \uppercase\expandafter{\romannumeral2}.4.2: Alice and Bob's Measurement.} Same as Protocol \uppercase\expandafter{\romannumeral1}.4.2.

\textbf{Step \uppercase\expandafter{\romannumeral2}.5: Basis Matching.} Same as Protocol \uppercase\expandafter{\romannumeral3}.5.

\textbf{Step \uppercase\expandafter{\romannumeral2}.6: Bit Mapping.} Same as Protocol \uppercase\expandafter{\romannumeral3}.6. 

\textbf{Step \uppercase\expandafter{\romannumeral2}.7: Parameter Estimation.} Alice and Bob use the total gain $Q'_{\mu \mu}$, QBER $E^{'Z}_{\mu \mu}$, the single photon rate $\Delta' _{11}$ and the single photon phase error rate $e^{X}_{11}$ to calculate the achievable secrecy rate $R'_{\mu \mu}$.

\textbf{Step \uppercase\expandafter{\romannumeral2}.8: Decoding.} Same as Protocol \uppercase\expandafter{\romannumeral3}.8.

Sending $\ket{\alpha e^{i\phi}}$ with a random phase $\phi$ is equivalent to sending $\ket{n}$ with probability $P_{\alpha^2}(n)$. Similarly, sending $\ket{\sqrt{2}\alpha e^{i\phi}}$ with a random phase is equivalent to sending $\ket{n}$ with probability $P_{2\alpha^2}(n)$. The probability is defined by $P_{\alpha^2}(n) = e^{-|\alpha|^2}\frac{(\alpha^2)^n}{n!}$ and $P_{2\alpha^2}(n) = e^{-2|\alpha|^2}\frac{(2\alpha^2)^n}{n!}$. When Alice and Bob each send out a single-photon state, the protocol \uppercase\expandafter{\romannumeral2} is equivalent to the protocol \uppercase\expandafter{\romannumeral1}. However, when they sent out multi-photon state, we have $\ket{\phi^X_{\theta_a}(\alpha)}\neq \ket{\phi^Z_{\theta_a}(\alpha)}$ and $\ket{\phi^X_{\theta_b}(\beta)}\neq \ket{\phi^Z_{\theta_b}(\beta)}$. Then, Eve can perform the PNS attack, which makes the protocol insecure. Thus the mutual information between Alice and Eve in this case is given by
\begin{equation}
    I(A:E)_{\text{multi}} = 1.
\end{equation}
Hence the total mutual information of Alice to Eve and Alice to Bob are given by
\begin{equation}
    \begin{aligned}
    I(A:E) &=\tilde{Q}_0 \times 0 + Q'_{11} h(e^{X}_{11}) + Q'_{\text{multi}}\times 1,\\
    I(A:B) &= Q'_{\mu \mu}[1-h(E_{\mu \mu}^{'Z})].
    \end{aligned}
\end{equation}
Therefore, the secrecy rate $R'$ of protocol \uppercase\expandafter{\romannumeral2} is given by
\begin{equation}
    \label{equ:R'}
    \begin{aligned}
        R' &= I(A:B) - I(A:E)\\
            &= Q'_{\mu \mu}\{\Delta'_{11}[1-h(e^X_{11})] - h(E^{'Z}_{\mu \mu})\}.
    \end{aligned}
\end{equation}
where $Q'_{\mu \mu}$ and $E^{'Z}_{\mu \mu}$ are the total gain and QBER when both of the intensities of the light sent by Alice and Bob are $\mu$, respectively, while $\Delta'_{11}=Q'_{11}/Q'_{\mu \mu}$ is the single-photon rate.

\subsection{Protocol \uppercase\expandafter{\romannumeral3}}
\label{app:protocol3}
\begin{figure*}[h]
    \centering
     \includegraphics[width=\textwidth]{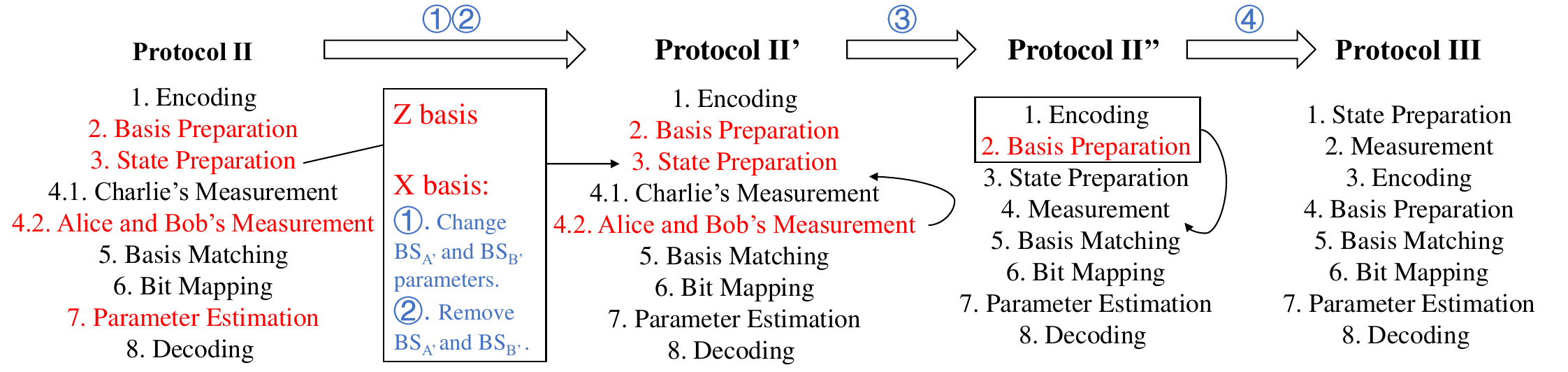}
     \caption{\label{fig:transmitters}\textbf{Transmitters used for equivalent transformation from Protocol \uppercase\expandafter{\romannumeral2} to Protocol \uppercase\expandafter{\romannumeral3}.} To transition from Protocol \uppercase\expandafter{\romannumeral2} to Protocol \uppercase\expandafter{\romannumeral3}, 4 equivalent transformations are required and are indicated by blue circled numbers. Protocol \uppercase\expandafter{\romannumeral2}' and Protocol \uppercase\expandafter{\romannumeral2}'' are intermediate protocols necessary for these transformations. The steps that differ in content from Protocol \uppercase\expandafter{\romannumeral3} are marked in red font.}
  \end{figure*}
From Protocol \uppercase\expandafter{\romannumeral2} to Protocol \uppercase\expandafter{\romannumeral3}, we need to perform four equivalent transformations as indicated by the blue circled numbers in Fig~\ref{fig:transmitters}. The transformation process is as follows.

\textcircled{1}. To estimate the single-photon rate and single-photon phase error rate, it is necessary to vary the intensity of the pulses of the two paths $i$ and $j$~\cite{Lo2005,Wang2005} in the X-basis. This can be achieved by adjusting the transmissivity and reflectivity of the beam splitters $\rm BS_{A'}$ and $\rm BS_{B'}$ so that the intensities satisfy $\alpha_i^2 + \alpha_j^2 = 2\alpha^2$, $\beta_i^2 + \beta_j^2 = 2\beta^2$, with $\alpha_i$, $\alpha_j$, $\beta_i$, and $\beta_j$ randomly distributed in $\{\sqrt{\nu}, \sqrt{\mu}\}$.

\textcircled{2}. Based on the previous statement, we analyze the quantum state after passing through the beam splitters $\rm BS_{A'}$ and $\rm BS_{B'}$. This state can be expressed as:
\begin{equation}
\begin{aligned}
\rho_{BS_{A'}} = \ket{\alpha_i e^{i\phi_a^i}} \bra{\alpha_i e^{i\phi_a^i}} \otimes \ket{\alpha_j e^{i\phi_a^j}} \bra{\alpha_j e^{i\phi_a^j}},\\
\rho_{BS_{B'}} = \ket{\beta_i e^{i\phi_b^i}} \bra{\beta_i e^{i\phi_b^i}} \otimes \ket{\beta_j e^{i\phi_b^j}} \bra{\beta_j e^{i\phi_b^j}}.
\end{aligned}
\end{equation}
This is equivalent to Alice and Bob randomly sending the quantum states $\ket{\alpha_i e^{i\phi_a^i}}$ and $\ket{\beta_i e^{i\phi_b^i}}$ through path $i$, and the quantum states $\ket{\alpha_j e^{i\phi_a^j}}$ and $\ket{\beta_j e^{i\phi_b^j}}$ through path $j$, where $\phi_a^j-\phi_a^i=\theta_a$ and $\phi_b^j-\phi_b^i=\theta_b$. Thus, it is possible to remove beam splitters $\rm BS_{A'}$ and $\rm BS_{B'}$ while still preserving the equivalence of the protocol.

\textcircled{3}. To further simplify the protocol from an entanglement-based version to a prepare-and-measure version, we can move Step \uppercase\expandafter{\romannumeral2}.4.2 before Step \uppercase\expandafter{\romannumeral2}.4.1 and merge it with Step \uppercase\expandafter{\romannumeral2}.3. This operation does not change the equivalence of the protocol since steps \uppercase\expandafter{\romannumeral2}.4.1 and \uppercase\expandafter{\romannumeral2}.4.2 are commutable. Consequently, Step \uppercase\expandafter{\romannumeral2}.3 becomes identical to Step \uppercase\expandafter{\romannumeral3}.1.

\textcircled{4}. Finally, it is necessary to assume a stable relative phase between events $i$ and $j$ in the X basis. If the time interval between events $i$ and $j$ is less than the maximum event interval $T$, this assumption is considered valid. Alice and Bob can use postselection to obtain $\theta_a$ and $\theta_b$, respectively. To accomplish this, Steps \uppercase\expandafter{\romannumeral2}.1 and \uppercase\expandafter{\romannumeral2}.2 can be moved after Steps \uppercase\expandafter{\romannumeral2}.3 and \uppercase\expandafter{\romannumeral2}.4, thus transforming the protocol into Protocol \uppercase\expandafter{\romannumeral3}.

Note that when the beam splitter $\rm BS_{A'}$ and $\rm BS_{B'}$ are removed, we need to consider paths $i$ and $j$ separately. Therefore, we need to derive the secrecy rate formula (\ref{equ:R'}) again. Here, we assume that Alice and Bob send pulses of $0$ intensity with a probability of $1/2$, and pulses of $\mu$ intensity with a probability of $1/2$. 

The average gain of each path is given by
\begin{equation}
        Q= \frac{1}{4}(Q_{\mu \mu}+Q_{\mu 0}+Q_{0 \mu}+Q_{00}),
\end{equation}
where $Q_{ab}(a,b\in \{0,\mu \})$ represents the gain when Alice sends the pulses of intensity $a$ while Bob sends the pulses of intensity $b$.

To calculate QBER and average single-photon bases rate, we need to consider paths $i$ and $j$ together. In our protocol, Alice and Bob use the Z basis to transmit messages. We consider the case where only one detector clicks for each of the two paths. The possible pulses' intensities combinations of Alice and Bob sent are 
\begin{equation}
    [a_ib_i,a_jb_j]\in \{ [0_i\mu_i,\mu_j0_j], [\mu_i0_i,0_j\mu_j], [0_i0_i,\mu_j\mu_j], [\mu_i\mu_i,0_j0_j] \}.
\end{equation}
The latter two cases cause QBER, which is given by
\begin{equation}
    E^Z= \frac{\frac{1}{16} (Q_{00}\cdot Q_{\mu \mu}+ Q_{\mu \mu}\cdot Q_{00})}{q\cdot Q^2},
\end{equation}
where $q$ represents the basis matching rate. And the average single-photon bases rate of each path is given by
\begin{equation}
    \Delta_1=   \frac{\frac{1}{16}\cdot 2\cdot P_{\mu}(1)^2(Y_1^2+ Y_2\cdot Y_0)}{q\cdot Q^2},
\end{equation}
where $Y_n$ is the yield of the case where there is $n$ photon in the channel.

Finally, the achievable secure rate $R$ is given by 
\begin{equation}
    R = p\cdot q\cdot Q \{ \Delta_1 [1-h(e^X_{11})]-fh(E^Z)\},
\end{equation}
where $p$ represents the basis preparation rate and $f$ represents the inefficiency function of error correction coding.

\section{Formulas of the performance analysis}
\label{app:performance}
In this section, we provide the formulas used for simulation. Our analysis is based on Appendix \ref{app:protocol3}. 

For basis preparation rate $p$, two single-photon interference events can prepare a basis, hence $p=1/2$.

For basis matching rate $k$, all of the possible pulses intensities' combinations of Alice and Bob sent are 
\begin{equation}
    [a_ib_i,a_jb_j]\in \{ [(0\backslash \mu)_i(0\backslash \mu)_i, (0\backslash \mu)_j(0\backslash \mu)_j] \}.
\end{equation}
Because Alice mostly prepared Z basis, we ignore the number of X basis. Thus, the effective combinations are:
\begin{equation}
    [a_ib_i,a_jb_j]\in \{ [0_i \mu_i,  \mu_j0_j],[ \mu_i0_i,  0_j\mu_j] \}.
\end{equation}
Hence $q\approx 2/16=1/8$.

For average gain $Q$ and QBER $E^Z$, we have~\cite{li2023one}
\begin{equation}
    \begin{aligned}
    Q_{\mu \mu} &= 1-e^{-2\eta \mu}+2p_de^{-2\eta \mu},\\
    Q_{\mu 0} &=Q_{0 \mu} =1-(1-2p_d)e^{-\eta \mu},\\
    Q_{00}&=2p_d(1-p_d),
    \end{aligned}
\end{equation}
where
\begin{equation}
    \begin{aligned}
    \eta&= \eta_d \sqrt{\eta_c} \\
     &=\eta_d 10^{-\frac{\zeta d}{20} },\\
    \end{aligned}
\end{equation}
where $\eta_d$ represents the detectors efficiecy,  $\zeta$ represents the fiber attenuation and $d$ represents the transmission distance.

We assume that there is no eavesdropper in the channel, then the single-photon yield $Y_1$ and the single-photon phase error rate $e_{11}$ are given by~\cite{li2023one,ma2012alternative}
\begin{equation}
    \begin{aligned}
    Y_1 &= 1-(1-2p_d)(1-\eta),\\
    Y_{11}&=(1-p_d)^2[\frac{\eta^2}{2}+p_d(4\eta -3\eta^2)+4p_d^2(1-\eta)^2 ],\\
    e_{11}&=e_0Y_{11}-(e_0-\delta )(1-p_d^2)\frac{\eta^2}{2},
    \end{aligned}
\end{equation}
where $e_0$ and $\delta $ represent the background error rate and misalignment, respectively.

\end{widetext}

\bibliographystyle{apsrev4-2}
\bibliography{moqsdc.bib}

\end{document}